\documentclass[aps,twocolumn,pra,superscriptaddress]{revtex4-1}

\usepackage[T1]{fontenc}
\usepackage[sc]{mathpazo}
\usepackage{amsmath}
\usepackage{amssymb}
\usepackage{enumerate}
\usepackage{amsthm}
\usepackage{amsfonts,mathrsfs}
\usepackage[style]{fncychap}
\usepackage{graphicx} 
\usepackage{geometry} 
\usepackage{eepic}
\usepackage{ifthen}
\usepackage{tikz}
\newboolean{ElectronicVersion}
\setboolean{ElectronicVersion}{true} 

\usepackage{hyperref}
\hypersetup{pdfpagemode=UseNone}

\newcommand{\setft}[1]{\mathrm{#1}}
\newcommand{\lin}[1]{\setft{L}\left(#1\right)}

\def\I{\mathbb{1}}

\newenvironment{mylist}[1]{\begin{list}{}{
    \setlength{\leftmargin}{#1}
    \setlength{\rightmargin}{0mm}
    \setlength{\labelsep}{2mm}
    \setlength{\labelwidth}{8mm}
    \setlength{\itemsep}{0mm}}}
    {\end{list}}

\def\ot{\otimes}

\newcommand{\out}[2]{| #1\rangle\langle #2 |}

\newcommand{\pa}[1]{(#1)}

\newcommand{\bra}[1]{\langle#1|}

\newcommand{\ket}[1]{|#1\rangle}

\DeclareMathOperator{\trace}{Tr}
\newcommand{\ptr}[2]{\trace_{#1}\pa{#2}}

\newcommand{\tr}[1]{\ptr{}{#1}}

\def\cD{\mathcal{D}}
\def\cH{\mathcal{H}}\def\cJ{\mathcal{J}}
\def\cM{\mathcal{M}}

\def\D{\textsf{D}}

\newtheorem{thrm}{Theorem}[section]
\newtheorem{lem}[thrm]{Lemma}
\newtheorem{prop}[thrm]{Proposition}

\theoremstyle{definition}

\newtheorem{remark}[thrm]{Remark}

\numberwithin{equation}{section}

\newcounter{questionnumber}

\begin{document}

\title{Logically reversible measurements: Construction and application}


 \author{Sunho Kim}
 \email{kimshanhao@126.com}
 \affiliation{School of Mathematical Sciences, Zhejiang University, Hangzhou 310027, PR~China}
 
 \author{Juncheng Wang}
 \email{dualinv@163.com}
 \affiliation{School of Mathematical Sciences, Zhejiang University, Hangzhou 310027, PR~China}

 \author{Asutosh Kumar}
 \email{usashrawan@gmail.com}
 \affiliation{The Institute of Mathematical Sciences, CIT Campus, Taramani, Chennai 600113, India}
 \affiliation{Homi Bhaba National Institute,  Anushaktinagar, Mumbai 400094, India}

 \author{Akihito Soeda}
 \email{soeda@phys.s.u-tokyo.ac.jp}
 \affiliation{Department of Physics, The University of Tokyo, Bunkyo Ku, Tokyo 113-0033, Japan}

 \author{Junde Wu}
 \email{Corresponding author: Junde Wu (wjd@zju.edu.cn)}
 \affiliation{School of Mathematical Sciences, Zhejiang University, Hangzhou 310027, PR~China}


\begin{abstract}
We show that for any von Neumann measurement, we can construct a logically reversible measurement such that
Shannon entropies and quantum discords induced by the two
measurements have compact connections. In particular, we prove that quantum discord for the logically reversible measurement is never less than that for the von Neumann measurement.
\end{abstract}

\maketitle


\section{Introduction}
Measurement, as envisaged, plays an inevitable role in quantum mechanics, and lies at the heart of ``interpretational problem'' of quantum mechanics. Nonetheless, different views of measurement almost universally agree on the measurement outcomes.
A quantum measurement is described in terms of a complete set of positive operators for the system to be measured. A few examples of quantum measurement are von Neumann measurement \cite{Holevo} which consists of orthogonal projectors, positive-operator-valued measure (POVM) \cite{Peres}, unitarily reversible measurement \cite{Mabuchi, Nielsen}, etc.  The most general type of measurement that can be performed on a quantum system is known as a generalized measurement \cite{Kraus, Gardiner}. Generalized measurements can be understood within the framework of quantum operations. Any measurement on a quantum state is inherently associated with wave function collapse and probability distribution. 
We recollect the necessary preliminaries briefly below.

{\em Quantum measurements.--}
Let $\cH$ be a finite dimensional complex Hilbert space, which
represents some quantum system. The set of quantum states $\rho$ on $\cH$ is denoted by $\D(\cH)$. A \emph{quantum measurement} on $\cH$ is a set $\Lambda
\equiv \{\Lambda_{x}\}_{x\in X} \subseteq \lin{\cH}$ of positive
operators indexed by $X$  and satisfies $\sum_{x} \Lambda_{x} = \mathbb{1}_{\cH}$.
Given a quantum state $\rho \in \D(\cH)$ and a quantum measurement
$\Lambda = \{\Lambda_x\}_{x\in X}$, then a probability distribution
$p = \{p(x)\}_{x\in X}$ is induced where $p(x) = Tr (\Lambda_{x}\rho)$ is the probability of the
outcome $x$ to occur. In this case, $\rho$ is transformed into the quantum state
$\rho_{x} = \frac{A_{x}\rho A_{x}}{p(x)}$, where $\Lambda_{x} = A_{x}^{2}$. If $\Pi = \{\Pi_{x}\}_{x\in X}$ is a set of orthogonal projectors,
then the measurement $\{\Pi_{x}\}_{x\in X}$ is said to be a
\emph{von Neumann measurement} \cite{Holevo}.
The celebrated Neumark extension theorem \cite{Spehner, Watrous} states that each quantum measurement can
be seen as a von Neumann measurement on a larger Hilbert space \cite{note1}.

We know that in a generalized measurement process, the input
state $\rho$ cannot always be retrieved with a nonzero success
probability by a ``reversing operation'' on the state $\rho_{x}$.
A measurement $\{\Lambda_{x}\}_{x\in X}$ is called \emph{logically reversible} \cite{Ueda} if the
premeasurement state $\rho$ of the measured system is uniquely determined
from the postmeasurement state $\rho_x$ and the outcome of the measurement.
Ueda {\em et al.} in Ref. \cite{Ueda} have shown that the measurement $\{\Lambda_{x}\}_{x\in X}$ is logically
reversible if and only if each measurement operator $\Lambda_{x}$ is
a reversible operator.
Moreover, if for each measurement operator $\Lambda_{x}$, there exists a unitary operator $U_x$ such that
\begin{eqnarray}
U_x\rho_xU_x^{\dagger} = \rho,
\end{eqnarray}
for each state $\rho$ whose support lies on a subspace $\cM$ of $\cH$, then $\{\Lambda_{x}\}_{x\in X}$ is called the {\em unitarily reversible} measurement \cite{Nielsen}.
It is clear that any von Neumann measurement $\{\Pi_{x}\}_{x\in X}$
is not logically reversible except $X$ has only a single element.
Note that in a logically reversible measuement, the system's information is preserved during the measurement process.
Thus, the reversibility of a measurement is related to the information gained from that measurement. 
Quantum teleportation \cite{Qtelep} can be seen as the problem of reversing a set of quantum operations \cite{Nielsen}.  

Suppose we are given a logically reversible measurement
$\Lambda_{u}=\{\Lambda_{u,x}\}_{x\in X}$. Since each measurement
operator $\Lambda_{u,x}$ is a positive (reversible) operator, then, by
the spectral decomposition theorem,
\begin{eqnarray}
\Lambda_{u,x} = \sum_{i\in\Sigma_{x}}a_x(i)\Pi_x(i),
\end{eqnarray}
where $\sum_{i\in\Sigma_{x}}\Pi_x(i) = \mathbb{1}_{\cH}$ and
$a_x(i)>0$ for any $i\in\Sigma_x$. In particular, if for all
$x\in X$ there exist subsets $\{i_s\}_{s=1}^{m_x} \subseteq
\Sigma_{x}$ such that $\sum_{s=1}^{m_x}\Pi_x(i_s)$ are the same
projector onto a subspace $\cM$ and $a_x(i_1) = \cdots =
a_x(i_{m_x})$, then the measurement $\Lambda_{u}$ is also a unitarily
reversible on the subspace $\cM$ \cite{Nielsen}.

The success probability $p_{s}$ of reversing, after the
measurement with result $x$, has the upper bound \cite{Korotkov, Jordan}
\begin{eqnarray}
p_{s} \leq \frac{\min_{i\in\Sigma_{x}}\{a_x(i)\}}{p_{u}(x)},
\end{eqnarray}
where $p_{u}(x) = Tr(\Lambda_{u,x}\rho)$.
If we define the \emph{total success probability} $p_{s}^{total}$ of reversing as
\begin{eqnarray}
p_{s}^{total} = \sum_{x\in X}p_{u}(x)p_{s},
\end{eqnarray}
then
\begin{eqnarray}\label{eq:upper bound}
p_{s}^{total} \leq \sum_{x\in X}\min_{i\in\Sigma_{x}}\{a_x(i)\}.
\end{eqnarray}
Note that the above bound 
is independent of the quantum state $\rho$.

{\em Shannon and von Neumann entropies.--}
A classical state is described by a probability distribution.
\emph{Shannon entropy} $H(p)$, for the probability distribution $p = \{p(x)\}_{x\in X}$, is defined by \cite{Shannon}
\begin{eqnarray}
H(p) = -\sum_{x\in X}p(x)\log_2p(x).
\end{eqnarray}
For a quantum state $\rho\in\D(\cH)$, the quantum analog of
Shannon entropy is \emph{von Neumann entropy}, and is given by
\begin{eqnarray}
S(\rho) = - Tr(\rho\log_2\rho).
\end{eqnarray}
An equivalent expression of $S(\rho)$ is \cite{Spehner},
\begin{eqnarray}
S(\rho) = \min_{\{\ket{\psi_{a}}, p_{a}\}} H(\{p_{a}\}),
\end{eqnarray}
where the minimum is taken over all pure state convex decompositions of $\rho$.
A decomposition minimizes $\{H(\{p_{a}\}):
\{\ket{\psi_{a}}, p_{a}\}\}$ if and only if it is a spectral
decomposition of $\rho$. For an arbitrary ensemble $\{\rho_{i}, \eta_{i}\}$, which forms a convex
decomposition of $\rho$, we have
\begin{eqnarray}\label{eq:entropy1}
S(\rho) \leq H(\{\eta_{i}\}) + \sum_{i}\eta_{i}S(\rho_{i})
\end{eqnarray}
The equality is achieved if and only if $\{\rho_{i}\}$ has mutual orthogonal supports.

{\em Quantum discord.--}
Let $\cH_{A}$ and $\cH_{B}$ be (the Hilbert spaces of) two quantum systems, $\rho_{AB}\in
\D(\cH_{A}\otimes\cH_{B})$ be a quantum state, $\rho_{A}$ and
$\rho_{B}$ be the reduced states of $\rho_{AB}$. In quantum
information theory, \emph{quantum mutual information}
\begin{eqnarray}
I_{A:B}(\rho_{AB}) = S(\rho_{A}) + S(\rho_{B}) - S(\rho_{AB}),
\end{eqnarray}
is regarded as a measure of the total correlation \cite{Groisman} between $\cH_{A}$
and $\cH_{B}$. 
With the quantum
conditional entropy, $S(\rho_{B}|\rho_{A}) = S(\rho_{AB}) -
S(\rho_{A})$, quantum mutual information becomes
\begin{eqnarray*}
 I_{A:B}(\rho_{AB}) = S(\rho_{B}) - S(\rho_{B}|\rho_{A}).
\end{eqnarray*}

Given a von Neumann measurement $\Pi^{A} = \{\Pi^{A}_{x}\}_{x\in X}$
on the quantum system $\cH_A$, let us define a conditional entropy
on the quantum system $\cH_B$ by $S_{B|A}(\rho_{AB}|\{\Pi^{A}_{x}\})
= \sum_{x}\eta_{x}S(\rho_{B|x}),$ where $\rho_{B|x} = \eta^{-1}_{x}Tr_{A}(\Pi^{A}_{x}\otimes
\mathbb{1}_{\cH_B}\rho_{AB})$ and $\eta_{x} = Tr(\Pi^{A}_{x}\otimes
\mathbb{1}_{\cH_B}\rho_{AB})$.
Moreover, we denote by
\begin{eqnarray}
\cJ_{B|A}^{vN}(\rho_{AB}) = S(\rho_{B}) -
\inf_{\Pi^{A}}\sum_{x}\eta_{x}S(\rho_{B|x}),
\end{eqnarray}
which is interpreted as a measure of classical correlation \cite{Henderson, Vedral} between
$\cH_A$ and $\cH_B$. In general, $I_{A:B}(\rho_{AB})$ and
$\cJ_{B|A}^{vN}(\rho_{AB})$ are different, and the difference between them
\begin{eqnarray}
\label{eq:qd-vn}
\cD_{A}^{vN}(\rho_{AB}) &=& I_{A:B}(\rho_{AB}) -
\cJ_{B|A}^{vN}(\rho_{AB}) \\
&=& S(\rho_{A}) - S(\rho_{AB}) +
\inf_{\Pi^{A}}\sum_{x}\eta_{x}S(\rho_{B|x}), \nonumber
\end{eqnarray} 
is called \emph{quantum discord}, which is interpreted as a measure of
quantum correlation \cite{Ollivier, Henderson, Vedral}.
It is an important {\em information-theoretic} measure of quantum correlation \cite{Modi}, beyond entanglement measures \cite{Horodecki}.

Moreover, if we replace the von Neumann measurement in (\ref{eq:qd-vn})
with the generalized quantum measurement $M^{A} =
\{M^{A}_{z}\}_{z\in Z}$ on $\cH_{A}$, then the general quantum
discord can be defined as follows:
\begin{eqnarray*}
\cD_{A}(\rho_{AB}) = S(\rho_{A}) - S(\rho_{AB}) +
\inf_{M^{A}}\sum_{z}\eta_{z}S(\rho_{B|z}),
\end{eqnarray*}
where
$\rho_{B|z} = \eta^{-1}_{z}Tr_{A}(\Lambda^{A}_{z}\otimes\mathbb{1}_{\cH_B}\rho_{AB})$ and $\eta_{z} = Tr(M^{A}_{z}\otimes\mathbb{1_{\cH_B}}\rho_{AB})$.
Clearly, $\cD_{A}(\rho_{AB}) \leq \cD_{A}^{vN}(\rho_{AB})$. 
Recall that, a \emph{purification} 
of $\rho\in\D(\cH_A)$ is any pure state $\out{\phi_{\rho}}{\phi_{\rho}}\in D(\cH_A\ot\cH_B)$
such that $Tr_{B}(\out{\phi_{\rho}}{\phi_{\rho}}) = \rho$.
It, then, follows from Neumark theorem and the additivity of von
Neumann entropy with respect to tensor products, that
\begin{eqnarray}\label{eq:discord}
\cD_{A}(\rho_{AB}) = \cD_{AE}^{vN}(\rho_{AB} \otimes
\out{\epsilon_{0}}{\epsilon_{0}}).
\end{eqnarray}

This paper is organized as follows. Section \ref{sec:rev-meas} deals with the construction of a class of logically reversible measurements based on a von Neumann measurement, and provides a relationship between Shannon entropies of the two measurements. Section \ref{sec:logical-qd} presents an inequality between quantum discords induced by the two measurements. Conclusion is presented in Section \ref{sec:conc}.

\section{Logically reversible measurements}
\label{sec:rev-meas}
In this section, we show that it is possible to construct a logically reversible measurement from any given von Neumann measurement, and establish a compact relation between Shannon entropies induced by the two measurements. 

Let $\rho\in\D(\cH)$ and $\Pi = \{\Pi_x\}_{x\in X}$ be a von Neumann
measurement with $|X| = n$. Now, based on $\Pi$ and any
$a\in(0,\frac{1}{n})$, we can construct the following logically reversible
measurement $\Lambda^{(a)}_{u} = \{\Lambda^{(a)}_{u,x}\}_{x\in X}$:
\begin{eqnarray}
\Lambda^{(a)}_{u,x} =\{1-(n-1)a\}\Pi_x + \sum_{y\neq x}a\Pi_y.
\end{eqnarray}
The probability distribution $p^{(a)}_{u} =
\{p^{(a)}_{u}(x)\}_{x\in X}$ is induced, and the probability
$p^{(a)}_{u}(x)$ of the classical outcome $x$ to occur is given by
\begin{eqnarray}
p^{(a)}_{u}(x) = Tr(\Lambda^{(a)}_{u,x}\rho) = (1-na)p(x) + a,
\end{eqnarray}
where $p(x)=Tr(\Pi_x\rho)$.
It is easy to show that the measurement $\Lambda^{(a)}_{u}$ is not unitarily reversible on any subspace $\cM$ with $\dim{\cM}\neq 1$ of $\cH$.
Note that the total success probability of reversing, after the
original von Neumann measurement $\Pi$, is zero. However, by
inequality (\ref{eq:upper bound}), the total success probability
$p_{s}^{total}$ of reversing, after the logically reversible measurement
$\Lambda^{(a)}_{u}$, has the nonzero upper bound
\begin{eqnarray}\label{eq:upper bound-2}
p_{s}^{total} \leq na.
\end{eqnarray}

Below, in Proposition~\ref{prop:Shannon entropy}, we give an important relationship between
Shannon entropies induced by the two measurements. We will adopt the notation, $H(p):=H(\{p(x)\})$.

\begin{prop}\label{prop:Shannon entropy}
For $\rho\in\D(\cH)$, and the logically reversible
measurement $\Lambda^{(a)}_{u} = \{\Lambda^{(a)}_{u,x}\}_{x\in X}$ which is induced by a von Neumann measurement $\Pi = \{\Pi_x\}_{x\in X}$ where $|X| = n$ and $a\in(0,\frac{1}{n})$, we observe the relation 
\begin{eqnarray*}
H(p_{u}^{(a)}) - n[\max\big\{f(a), f(1-na+a)\big\}] \leq H(p) \leq
H(p_{u}^{(a)}),
\end{eqnarray*}
where $p_{u}^{(a)}(x) = Tr(\Lambda_{u,x}^{(a)}\rho) = (1-na)p(x) + a$, $p(x)=Tr(\Pi_x\rho)$ and $f(x) = -x\log_2x$.
\end{prop}

\begin{proof}
Let us consider the following two sets:
\begin{eqnarray*}
A = \{x | p(x) \leq \frac{1}{n}\}, \ \ B = \{x | p(x) >
\frac{1}{n}\}.
\end{eqnarray*}
Note that for positive numbers $p\leq\frac{1}{n}$ and
$q>\frac{1}{n}$, we have
\begin{eqnarray*}
p \leq (1-na)p + a,\ \ \ q > (1-na)q + a.
\end{eqnarray*} 
Therefore,
\begin{eqnarray*}
0 \leq \sum_{x\in A}(p^{(a)}_{u}(x) - p(x)) = \sum_{x\in B}(p(x) -
p^{(a)}_{u}(x)).
\end{eqnarray*}

\begin{figure}[htb]
\includegraphics[scale=0.3]{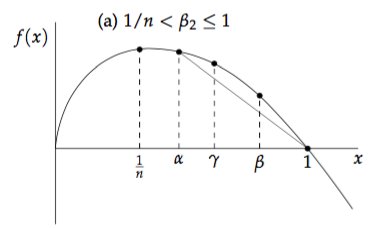}
\includegraphics[scale=0.3]{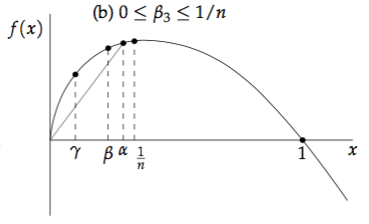}
\caption{$f(\alpha) \leq f(\beta) + f(\gamma)$.}
\label{fig:fig1}
\end{figure}

Let us denote by $\alpha_1 = \max_{x\in
A}\{p^{(a)}_{u}(x)\}$, $ \beta_1 = \min_{x\in
B}\{p^{(a)}_{u}(x)\}$, and $\gamma_1 = \sum_{x\in
A}(p^{(a)}_{u}(x) - p(x)) = \sum_{x\in B}(p(x) - p^{(a)}_{u}(x))$. 
Also, let $f(x) = -x\log_2x$. Then, for any $0 < p \leq q \leq 1$, we have $f'(p) \geq f'(q)$.
Hence,
\begin{eqnarray*}
&&\sum_{x\in A} [-p^{(a)}_{u}(x)\log_{2}p^{(a)}_{u}(x) + p(x)\log_{2}p(x)] \nonumber \\
&\geq& \gamma_1 f'(\alpha_1) \geq \gamma_1 f'(\beta_1) \nonumber \\
&\geq& \sum_{x\in B} [-p(x)\log_{2}p(x) + p^{(a)}_{u}(x)\log_{2}p^{(a)}_{u}(x)].
\end{eqnarray*}
Thus, we obtain one-half of Proposition~\ref{prop:Shannon entropy}.
\begin{eqnarray}
\label{eq:undoable measurement-1}
H(p_{u}^{(a)}) \geq H(p).
\end{eqnarray}

\begin{widetext}
\noindent To prove the other half of Proposition~\ref{prop:Shannon entropy}, we consider two cases separately.\\
{\em Case 1:}
If $p(x)>\frac{1}{n}$, let $\alpha_2 = (1-na)p(x)+a \left(= p^{(a)}_{u}(x)\right)$, 
$\beta_2 = p(x)$ and $\gamma_2 = 1-na+a$. Then  $\frac{1}{n} <
\alpha_2, \beta_2, \gamma_2 \leq 1$, $0 < \gamma_2 - \alpha_2 \leq 1
- \beta_2,$ and $f(\alpha_2) \leq f(\beta_2) + f(\gamma_2)$ (see Fig.~\ref{fig:fig1}(a)). Hence,
\begin{eqnarray}
\label{eq:case1}
-p_{u}^{(a)}(x)\log_2p_{u}^{(a)}(x) 
\leq -p(x)\log_2 p(x) 
 -  (1-na+a)\log_2(1-na+a).
\end{eqnarray}

\noindent {\em Case 2:}
If $p(x)\leq\frac{1}{n},$ let $\alpha_3 = (1-na)p(x)+a$, $ \beta_3
= p(x)$, and $\gamma_3 = a$. Then $0 < \alpha_3, \beta_3,
\gamma_3 \leq \frac{1}{n}$,  $\alpha_3 \leq \beta_3 + \gamma_3$, and $f(\alpha_3) \leq f(\beta_3) + f(\gamma_3)$
(see Fig.~\ref{fig:fig1}(b)). Hence,
\begin{eqnarray}
\label{eq:case2}
-p_{u}^{(a)}(x)\log_2p_{u}^{(a)}(x)
\leq -p(x)\log_2p(x) - a\log_2a.
\end{eqnarray}

\noindent Now, summing (\ref{eq:case1}) and (\ref{eq:case2}) over allowed probabilities and adding them, we obtain
\begin{eqnarray}
\label{eq:undoable measurement-2}
H(p_{u}^{(a)}) - n\Big[\max\big\{f(a), f(1-na+a)\big\}\Big] \leq H(p).
\end{eqnarray} 
\end{widetext}
Combining (\ref{eq:undoable measurement-1}) and (\ref{eq:undoable measurement-2}), the proposition is proved.
\end{proof}

\begin{remark}
Note that
$\lim_{a\rightarrow 0}\max\big\{f(a), f(1-na+a)\big\}=0$.
So, it follows from Proposition \ref{prop:Shannon entropy} that
$\lim_{a\rightarrow 0}H(p_{u}^{(a)}) = H(p)$.
This is expected because when $a\rightarrow 0$, $\Lambda^{(a)}_u \rightarrow \Pi$.
\end{remark}

\section{Quantum discord for logically reversible measurements}
\label{sec:logical-qd}

In this section, we study quantum discord with respect to logically reversible measurement $\Lambda^{(a,A)}_{u}$ on $\cH_{A}$, where dim $\cH_{A} = n$ and $a\in(0,\frac{1}{n})$.
Quantum discord of state $\rho_{AB}\in\D(\cH_A\otimes\cH_B)$ for the logically reversible measurement is defined by
\begin{eqnarray*}
\cD^{(a)}_{u,A}(\rho_{AB}) = I_{A:B}(\rho_{AB}) - \cJ_{u,B|A}^{(a)}(\rho_{AB}),
\end{eqnarray*}
where 
\begin{eqnarray*}
\cJ_{u,B|A}^{(a)}(\rho_{AB}) &=& S(\rho_{B}) - \inf_{\Lambda^{(a,A)}_{u}} \sum_{x}\eta_{u,x}S(\rho^{(a)}_{u,B|x}), \nonumber \\
\rho^{(a)}_{u,B|x}& =& \eta^{-1}_{u,x}Tr_{A}(\Lambda^{(a,A)}_{u,x}\otimes \mathbb{1}_{B}\rho_{AB}), \nonumber \\
\eta_{u,x} &=&Tr(\Lambda^{(a,A)}_{u,x}\otimes\mathbb{1}_{B}\rho_{AB}). \nonumber \\ 
\end{eqnarray*}
In the following, we establish an important relation between quantum discord for von Neumann measurement, $\cD^{vN}_{A}(\rho_{AB})$, and quantum discord for logically reversible measurement, $\cD^{(a)}_{u,A}(\rho_{AB})$. For this we need Lemma~\ref{lem:entropy}.

\begin{lem}\label{lem:entropy} 
Let $\rho,~\rho_{1},~\rho_{2}\in\D(\cH_A)$, $p_{0} + p_{1} + p_{2} = 1$, $\rho = (p_{0}+p_{1})\rho_{1} + p_{2}\rho_{2}$, and 
$H_0(r) = - r\log_{2}r - (1-r)\log_{2}(1-r)$ \ for any $r\in[0,1]$. Then,
\begin{eqnarray*}
\label{eq:lemma}
S(\rho) &\leq& p_{0}S(\rho_{1}) + (p_{1}+p_{2})S\left(\frac{p_{1}}{p_{1}+p_{2}}\rho_{1}+\frac{p_{2}}{p_{1}+p_{2}}\rho_{2}\right) \nonumber \\
& -& (p_{1}+p_{2})H_0\left(\frac{p_{1}}{p_{1}+p_{2}}\right) + H_0(p_{2}).
\end{eqnarray*}
\end{lem}

\begin{proof}
Let us introduce two quantum systems $\cH_B$ and $\cH_C$, and
construct a quantum state
$\rho_{ABC}\in\D(\cH_A\otimes\cH_B\otimes\cH_C)$ as $\rho_{ABC} =
p_{0}\rho_{1}\otimes\ket{0}_{B}\bra{0}\otimes\ket{0}_{C}\bra{0} +
p_{1}\rho_{1}\otimes\ket{0}_{B}\bra{0}\otimes\ket{1}_{C}\bra{1} +
p_{2}\rho_{2}\otimes\ket{1}_{B}\bra{1}\otimes\ket{1}_{C}\bra{1}$.
Then, we have $S(\rho_{A}) = S(\rho)$, $S(\rho_{AB}) = H_0(p_{2}) + (p_{0}+p_{1})S(\rho_{1}) + p_{2}S(\rho_{2})$, 
$S(\rho_{AC}) = H_0(p_{0}) + p_{0}S(\rho_{1}) +
(p_{1}+p_{2})S\left(\frac{p_{1}}{p_{1}+p_{2}}\rho_{1}+\frac{p_{2}}{p_{1}+p_{2}}\rho_{2}\right)$, and $S(\rho_{ABC}) = H(p) + (p_{0}+p_{1})S(\rho_{1}) + p_{2}S(\rho_{2})$, 
where the probability distribution $p = (p_{0},p_{1},p_{2})$. 
Now, exploiting the strong subadditivity of von Neumann entropy \cite{Watrous}, $S(\rho_{ABC}) + S(\rho_{A}) \leq S(\rho_{AB}) + S(\rho_{AC})$, and simplifying we obtain the desired result.
\end{proof}

\begin{thrm}\label{th:main result} For $\rho_{AB}\in\D(\cH_A\otimes\cH_B)$ with $\dim{\cH_A} = n$, $a\in(0,\frac{1}{n})$, and the probability distribution $p_{n,a} = \{(1-(n-1)a),\overbrace{a,\cdots,a}^{n-1}\}$, we have
\begin{eqnarray*}
&&\cD^{(a)}_{u,A}(\rho_{AB}) -
\frac{na\cJ_{u,B|A}^{(a)}(\rho_{AB})}{1-na} - H(p_{n,a}) \nonumber \\
& \leq & \cD_{A}^{vN}(\rho_{AB}) \leq \cD^{(a)}_{u,A}(\rho_{AB}) -
\frac{na\cJ_{u,B|A}^{(a)}(\rho_{AB})}{1-na}.
\end{eqnarray*}
\end{thrm}

\begin{proof}
Let $\Lambda^{(a,A)}_{u} = \{\Lambda^{(a,A)}_{u,x}\}_{x\in X}$ be the logically reversible measurement induced by
von Neumann measurement $\Pi^{A} = \{\Pi^{A}_x\}_{x\in X}$.That is,
\begin{eqnarray*}
\Lambda^{(a,A)}_{u,x} =(1-(k-1)a)\Pi^{A}_x + \sum_{y\neq x}a\Pi^{A}_y,
\end{eqnarray*}
where $k = |X|$ and $a\in(0,\frac{1}{k})$. Then the conditional state is
\begin{eqnarray*}
\rho^{(a)}_{u,B|x}&=&\eta^{-1}_{u,x}Tr_{A}(\Lambda^{(a,A)}_{u,x}\otimes
\mathbb{1}_{B}\rho_{AB}) \\
&=&\frac{(1-ka)\eta_{x}}{\eta_{u,x}}\rho_{B|x} +
\frac{a}{\eta_{u,x}}\rho_{B},
\end{eqnarray*}
where $\eta_{u,x} = (1-ka)\eta_{x}+a$, $\eta_{x} = Tr(\Pi^{A}_{x}\otimes\mathbb{1}_{B}\rho_{AB})$,
 and $\rho_{B} = Tr_{A}(\rho_{AB})$.
Thus, by the concavity of von Neumann entropy, we have
\begin{eqnarray*}
\frac{(1-ka)\eta_{x}}{\eta_{u,x}}S(\rho_{B|x}) + \frac{a}{\eta_{u,x}}S(\rho_{B}) \leq S(\rho^{(a)}_{u,B|x}),
\end{eqnarray*}
implying
\begin{eqnarray}
(1-ka) \sum_{x\in X}\eta_{x}S(\rho_{B|x}) + kaS(\rho_{B})
&\leq& \sum_{x\in X}\eta_{u,x}S(\rho^{(a)}_{u,B|x}).\nonumber \\
\label{eq:qd1}
\end{eqnarray}
Because $\inf_{\Pi^{A}} \sum_{x}\eta_{x}S(\rho_{B|x})$ is
achieved on rank-one projectors, $k = |X| = \dim \cH_A = n$.
Therefore, using (\ref{eq:qd1}), we have
\begin{eqnarray}\label{eq:undoable discord-1}
\frac{\cJ_{u,B|A}^{(a)}(\rho_{AB})}{1-na} \leq \cJ_{B|A}^{vN}(\rho_{AB}).
\end{eqnarray}

Besides, if we denote by
$\rho_{B|X\setminus\{x\}}=\frac{\rho_{B}-\eta_{x}\rho_{B|x}}{1-\eta_{x}}$, then
\begin{eqnarray*}
\rho^{(a)}_{u,B|x} = \frac{(1-(n-1)a)\eta_{x}}{\eta_{u,x}}\rho_{B|x}
+ \frac{a(1-\eta_{x})}{\eta_{u,x}}\rho_{B|X\setminus\{x\}},
\end{eqnarray*}
and
$\rho_{B} = \eta_{x}\rho_{B|x}+(1-\eta_{x})\rho_{B|X\setminus\{x\}}$.

\begin{widetext}
Let $p_0=\frac{(1-na)\eta_{x}}{\eta_{u,x}},
p_1=\frac{a\eta_{x}}{\eta_{u,x}},
p_2=\frac{a(1-\eta_{x})}{\eta_{u,x}}, \rho_1=\rho_{B|x},
\rho_2=\rho_{B|X\setminus\{x\}}$. Then, by Lemma \ref{lem:entropy},
we have
\begin{eqnarray*}
S(\rho^{(a)}_{u,B|x})&\leq&\frac{(1-na)\eta_{x}}{\eta_{u,x}}S(\rho_{B|x})+
\frac{a}{\eta_{u,x}}S(\rho_{B})-\frac{a}{\eta_{u,x}}H_0(\eta_{x}) +
H_0(\frac{a(1-\eta_{x})}{\eta_{u,x}}).
\end{eqnarray*}
After simple algebra, and using (\ref{eq:undoable measurement-1}), we obtain
\begin{eqnarray*}
\sum_{x\in X}\eta_{u,x}S(\rho^{(a)}_{u,B|x}) \leq
(1-na) \sum_{x\in X}\eta_{x}S(\rho_{B|x}) + naS(\rho_{B}) + H(p_{n,a}) - naH(\eta_u),
\end{eqnarray*}
where $H(p_{n,a}) = - (1-(n-1)a)\log_{2}(1-(n-1)a) - (n-1)a\log_{2}a$ and $H(\eta_u) = - \sum_{x\in X}\eta_{u,x}\log_{2}\eta_{u,x}$.
Also, since $a\leq \eta_{u,x} \leq 1-(n-1)a$ for all $x$, we have $H(p_{n,a}) \leq H(\eta_u)$. Therefore,
\begin{eqnarray}\label{eq:undoable discord-2}
\sum_{x\in X}\eta_{u,x}S(\rho^{(a)}_{u,B|x}) \leq (1-na) \sum_{x\in X}\eta_{x}S(\rho_{B|x}) + naS(\rho_{B}) + (1-na)H(p_{n,a}),
\end{eqnarray}
and
\begin{eqnarray}
\label{eq:qd2}
\cJ_{B|A}^{vN}(\rho_{AB})=S(\rho_{B}) - \inf_{\Pi^{A}} \sum_{x\in X}\eta_{x}S(\rho_{B|x})
\leq \frac{\cJ_{u,B|A}^{(a)}(\rho_{AB})}{1-na} + H(p_{n,a}).
\end{eqnarray}
\end{widetext}

Now, 
\begin{eqnarray*}
&&\cD^{(a)}_{u,A}(\rho_{AB}) - \frac{na\cJ_{u,B|A}^{(a)}(\rho_{AB})}{1-na} - H(p_{n,a})\\
&=&I_{A:B}(\rho_{AB}) - \frac{\cJ_{u,B|A}^{(a)}(\rho_{AB})}{1-na} - H(p_{n,a}) \\
&\leq&I_{A:B}(\rho_{AB}) - \cJ_{B|A}^{vN}(\rho_{AB}) = \cD_{A}^{vN}(\rho_{AB})\\
&\leq& I_{A:B}(\rho_{AB}) - \frac{\cJ_{u,B|A}^{(a)}(\rho_{AB})}{1-na}\\
&=&\cD^{(a)}_{u,A}(\rho_{AB}) - \frac{na\cJ_{u,B|A}^{(a)}(\rho_{AB})}{1-na},
\end{eqnarray*}
where the first inequality is due to (\ref{eq:qd2}), and the second inequality follows from (\ref{eq:undoable discord-1}). Hence, the proof.
\end{proof}

\begin{remark} 
Note that $\cJ_{u,B|A}^{(a)}(\rho_{AB}) > 0$, and from Theorem~\ref{th:main result}, we have
\begin{eqnarray*}
\cD^{(a)}_{u,A}(\rho_{AB}) - \frac{na\cJ_{u,B|A}^{(a)}(\rho_{AB})}{1-na} \geq \cD_{A}^{vN}(\rho_{AB}),
\end{eqnarray*}
where $a \in (0,1/n)$. Hence,
\begin{eqnarray*}
\cD^{(a)}_{u,A}(\rho_{AB}) - \cD_{A}^{vN}(\rho_{AB})\ \geq \
\frac{na\cJ_{u,B|A}^{(a)}(\rho_{AB})}{1-na} > 0.
\end{eqnarray*}
Thus, with the logically reversible measurement, one can extract more quantum discord than with the von Neumann measurement. See also \cite{note2}.
\end{remark}

\section{Conclusion}
\label{sec:conc}

In summary, we constructed the logically reversible measurement based on the von Neumann measurement. We then established relationships for Shannon entropies, and quantum discords with respect to these two measurements. In particular, we showed that quantum discord for the logically reversible measurement exceeds that for the von Neumann measurement.

\subsection*{Acknowledgements} 
AK acknowledges the research fellowship of Department of Atomic Energy, Government of India.
This  project is supported by National Natural Science Foundation of China (11171301, 11571307)
and by the Doctoral Programs Foundation of the Ministry of Education of China (20120101110050).




\begin{thebibliography}{99}
\bibitem{note2} Moreover, by Eq. (\ref{eq:discord}), we get the following important estimate of the general quantum discord $\cD_{A}(\rho_{AB})$:
$\cD^{(a)}_{u,AE}(\rho_{AB}\otimes \out{\epsilon_{0}}{\epsilon_{0}}) - \frac{na\cJ_{u,B|AE}^{(a)}(\rho_{AB}\otimes \out{\epsilon_{0}}{\epsilon_{0}})}{1-na}\ -\ H(p_{n,a})
\leq \cD_{A}(\rho_{AB})\\ 
\leq \ \cD_{AE}^{vN}(\rho_{AB}\otimes
\out{\epsilon_{0}}{\epsilon_{0}}) -
\frac{na\cJ_{u,B|AE}^{(a)}(\rho_{AB}\otimes
\out{\epsilon_{0}}{\epsilon_{0}})}{1-na}$.
\end{thebibliography}
\end{document}